 \documentclass{sig-alternate}
\usepackage{stmaryrd}
\usepackage{listings}
\usepackage{multirow}
\usepackage{array}
\usepackage{subfig}
\usepackage{algorithmic}
\usepackage{algorithm}
\usepackage{amsmath}
\usepackage{amssymb}
\usepackage{xspace}
\usepackage{graphicx}
\usepackage{siunitx}
\usepackage{qtree}
\usepackage[svgnames,table]{xcolor}
\usepackage{tikz}
\usepackage{enumitem}
\usetikzlibrary{decorations.markings,shapes,arrows,positioning}

\newcommand\code[1]{\textsf{\small #1}}

\newcommand\tuple[1]{\langle #1 \rangle}

\newcommand\tool{\text{DIME*}\xspace}
\newcommand\oldtool{\text{DIME}\xspace}
\newcommand\cond{\emph{Trace Conditional}\xspace}
\newcommand\tv{\emph{Trace Version}\xspace}
\newcommand\tvc{\emph{Trace Version Conditional}\xspace}
\newcommand\enable{\emph{DBI-enabled}\xspace}
\newcommand\disable{\emph{DBI-disabled}\xspace}
\newcommand\vins{\text{V\_INSTRUMENT}\xspace}
\newcommand\vbase{\text{V\_BASE}\xspace}

\makeatletter
\def\@copyrightspace{\relax}
\makeatother

\begin{document}

\title{Redundancy Suppression In Time-Aware Dynamic Binary Instrumentation}


\numberofauthors{3}
\author{
\alignauthor
Pansy Arafa
\\ 
      \affaddr{University of Waterloo}\\
      \email{parafa@uwaterloo.ca}
 \alignauthor
 Hany Kashif
 \\ 
      \affaddr{University of Waterloo}\\
      \email{hkashif@uwaterloo.ca}
 \alignauthor
 Sebastian Fischmeister
 \\ 
      \affaddr{University of Waterloo}\\
      \email{sfischme@uwaterloo.ca}
}
\date{}

\maketitle

\begin{abstract}

Software tracing techniques are well-established and used by instrumentation tools to extract run-time information for program analysis and debugging.
Dynamic binary instrumentation as one tool instruments program binaries to extract information.
Unfortunately, instrumentation causes perturbation that is unacceptable for time-sensitive applications.
Consequently we developed \tool, a tool for dynamic binary instrumentation that considers timing constraints.
\tool uses Pin and a rate-based server approach to extract information only as long as user-specified constraints are maintained.
Due to the large amount of redundancies in program traces, \tool reduces the instrumentation overhead by one to three orders of magnitude compared to native Pin while extracting up to 99\% of the information.
We instrument VLC and PostgreSQL to demonstrate the usability of \tool.
\end{abstract}
%

\section{Introduction}
\label{sec:intro}

Program profiling is essential for analyzing program performance and understanding run-time behavior~\cite{Upton2011, Bock2011,Rico2011}.
A program profiler extracts program information during execution; such as memory access patterns, code coverage, register usage, and dynamic call context trees.
Other examples include collection of run-time statistics on instruction usage, or cache hits and misses.
In order to extract this run-time information, the profiler instruments the program.
Instrumentation naturally causes perturbation to the program under analysis.
Such perturbation can result in erroneous performance-analysis data or inaccurate program profile.

Instrumentation of performance-sensitive applications such as video players must not only preserve functional correctness, but also performance constraints.
Thus, time-sensitive applications require specialized program-profiling tools in order to honor their timing constraints.
Generally, there exist two instrumentation approaches; hardware based and software based.
Hardware-based tracing methods~\cite{Moore2003,Omre2008} cause significant perturbation to the traced program~\cite{Mytkowicz2008}.
Also, these methods collect low-level data and, hence, require higher-level support to provide traces at a higher-level of abstraction~\cite{Mork,Moseley2007}.
Software-based instrumentation methods~\cite{Mellor-Crummey1989,Kim2004} insert code in the original program, either statically or dynamically, to enable tracing, which results in modifying the program's timing behavior.
Consequently, recent line of research focuses on time-aware instrumentation techniques that respect the timing constraints of the program~\cite{Fischmeister2010a,Kashif2012,Kashif2013,Arafa2013}.

Time-aware instrumentation extracts program information while preserving both functional and timing properties.
Fischmeister and Lam~\cite{Fischmeister2010a} propose a time-aware static instrumentation technique that preserves the worst-case execution time (WCET) of the instrumented program.
Kashif et al.~\cite{Kashif2012} apply program transformation techniques to increase the effectiveness of time-aware instrumentation.
In~\cite{Kashif2013}, Kashif et al. introduce INSTEP; a static instrumentation framework for preserving extra-functional properties.
All these approaches are impractical for large code bases with library dependencies as they require static analysis and worst-case execution time analysis.
They also do not support multi-threaded applications.
In~\cite{Arafa2013}, Arafa et al. propose \oldtool; a time-aware dynamic binary instrumentation framework using rate-based resource allocation.
\oldtool limits the instrumentation time to a predefined budget per each time period.
If the instrumentation consumes the budget before the time period ends, \oldtool will disable the instrumentation till the beginning of the next time period.
While \oldtool reduces overhead, it only extracts partial (incomplete) traces due to disabling instrumentation on budget consumption.

In this paper, we introduce \tool, which offers the same advantages as the original version \oldtool, but provides higher instrumentation coverage.
The idea is to attempt to trace only untraced information in successive runs of the program under analysis.
Even with multiple runs, running a program multiple times on top of \tool is less time-consuming than native Pin.
We investigate different implementations and compare them qualitatively.
We quantitatively evaluate the performance of \tool compared to native Pin using the SPEC benchmarks~\cite{spec}.
Our results show that \tool extracts up to 99\% of the tracing information while reducing the overhead of native Pin by up to three orders of magnitude.
For instance, \code{dealII}, a SPEC benchmark~\cite{spec}, originally runs for four minutes without any instrumentation.
Extracting the branch profile of \code{dealII} using native Pin takes nine days of CPU time, while it only takes a maximum of 30 minutes using \tool.
We also apply our approach to VLC~\cite{VLC} and PostgreSQL~\cite{postgres} as case studies to demonstrate the scalability and applicability of \tool.



\section{Background}
\label{sec:back}

This section provides a basic overview of time-aware instrumentation, dynamic binary instrumentation, and the basics of \oldtool.

\subsection{Time-aware Instrumentation}
\label{sec:taware}
Instrumentation is the process of inserting extra code inside a program to extract information during execution.
Instrumentation can be static or dynamic.
Static instrumentation means inserting the instrumentation code before running the program, whereas dynamic techniques inserts the instrumentation code during the program execution.
Time-aware instrumentation~\cite{Fischmeister2010a,Kashif2012,Kashif2013,Arafa2013} is a mechanism that preserves functional correctness and timing properties when instrumenting programs.
The static technique in~\cite{Fischmeister2010a} instruments a program at code locations that do not modify the program's WCET and at the same time preserves the program's original behavior.
Static time-aware instrumentation shifts the program's execution time profile towards its deadline.
The authors in~\cite{Kashif2012} apply code transformation techniques to the program under analysis to increase instrumentation coverage.
They duplicate or create basic blocks in the program to increase the locations at which instrumentation code can be inserted while preserving timing constraints.
The authors in~\cite{Kashif2013} introduce INSTEP; an instrumentation framework for preserving multiple competing extra-functional properties.
INSTEP uses cost models and constraints of the extra-functional properties in addition to the user's instrumentation intent to transform the input program into an instrumented program that respects the specified constraints.
All the mentioned work on time-aware instrumentation is based on static source-code instrumentation techniques.
Static source-code instrumentation requires performing WCET analysis of the input program to guide the placement of instrumentation code inside the input program.
It also needs WCET analysis after instrumenting the program to validate that timing constraints are met.
Static instrumentation techniques are sound and effective, but the need for running WCET analysis before and after instrumentation reduces the applicability to only hard real-time applications where WCET analysis is common.
Additionally, static instrumentation requires the availability of the source code including all library dependencies.
For example, the VLC media player~\cite{VLC} has approximately \num{600000} lines of code and uses libraries with more than three million lines of code.
Thus, it is impractical to statically analyze the source code of a multi-threaded application like VLC along with its library dependencies.
In~\cite{Arafa2013}, Arafa et al. propose \oldtool; a time-aware dynamic binary instrumentation framework.

\subsection{Dynamic Binary Instrumentation}
\label{sec:dbi}
Dynamic binary instrumentation (DBI) instruments the program's binary  during execution.
Unlike static instrumentation, DBI does not require preprocessing of the program under analysis.
On the other hand, DBI incurs higher run-time overhead compared to static instrumentation since DBI decides the placement of instrumentation code at runtime.

Pin~\cite{Luk2005} is a DBI instrumentation framework that provides a cross-platform API for building program-profiling tools.
Pin has lower run-time overhead than that of the other DBI frameworks like DynamoRio~\cite{Bruening2003}, and Valgrind~\cite{Nethercote:2007}.
Pin is easily extensible and transparent i.e., it maintains the same instruction and data addresses, and the same register and memory values compared to the original program.
Pin can trace statically unknown indirect-jump targets, dynamically generated code, and dynamically loaded libraries.
It can also handle mixed code and data, and variable-length instructions.
To build an analysis tool using Pin (pintool), two types of routines should be implemented. 
The analysis routine contains the code to be inserted in the program during execution, whereas the instrumentation routine decides on the locations of inserting the analysis-routine calls.
The analysis routine is the main source of Pin's overhead which varies according to the invocation frequency of the analysis routines and their complexity.
On the other hand, the dynamic compilation and the execution of the instrumentation routine represent a minor source of run-time overhead.
Pin is known for its efficiency~\cite{Luk2005, Arafa2013}; it uses a just-in-time (JIT) compiler to insert and optimize instrumentation code. 
The unit of compilation is the trace; a straight-line sequence of instructions that have a single entry point and may have multiple exits.
When the program starts execution, Pin compiles the first trace and generates a modified one which is almost identical to the original.
The modified trace enables Pin to regain control when needed. 
Pin transfers control to the generated trace, then regains control when a branch exits the trace. 
Afterwards, Pin compiles the new trace and continues execution. 
Whenever the JIT compiler fetches some code to compile, the pintool is allowed to instrument the code before compilation. 
Pin saves the compiled code and its instrumentation in a cache in case it gets re-executed~\cite{Uh2007, Ruiz2008, Luk2005}.
Pin supports different granularities for both the instrumentation routine and the analysis routine, e.g., trace, routine, and  instruction granularities.
The instrumentation-routine granularity defines when Pin should execute the instrumentation routine.
Similarly, the granularity of the analysis routine tells Pin where to insert the analysis-routine calls.
\oldtool~\cite{Arafa2013} is implemented as an extension to Pin~\cite{Luk2005}.

\subsection{\oldtool}
\label{sec:dime}
\oldtool~\cite{Arafa2013} is a dynamic binary time-aware instrumentation tool that respects the timing properties of the program.
\oldtool, as a tool for instrumenting soft real-time applications, is practical, scalable, and supports multi-threaded applications. 
It guarantees less run-time overhead compared to Pin especially for the profiling tools with heavy-weight analysis routines.
\oldtool uses rate-based resource allocation to limit the instrumentation time to a pre-specified budget $B$ per time period $T$.
Instrumentation is enabled (i.e. allowed to execute) for a total of $t_{ins}$ time units in every time period $T$.
The total instrumentation time $t_{ins}$ per period $T$ should not exceed the instrumentation budget $B$.
If the budget is fully consumed before the end of the time period $T$, \oldtool will disable instrumentation.
At the beginning of the next period $T$, the budget resets to $B$ time units and the instrumentation is re-enabled. 
This process repeats until the program terminates.
Specifically, \oldtool limits the execution time of the analysis routine to the budget $B$~\cite{Arafa2013}.
The reason is that the analysis routine is the main source of run-time overhead, and the overhead of the instrumentation routine is negligible~\cite{Luk2005}.

\begin{figure}[htb]
  \centering \includegraphics[width=.9\linewidth]{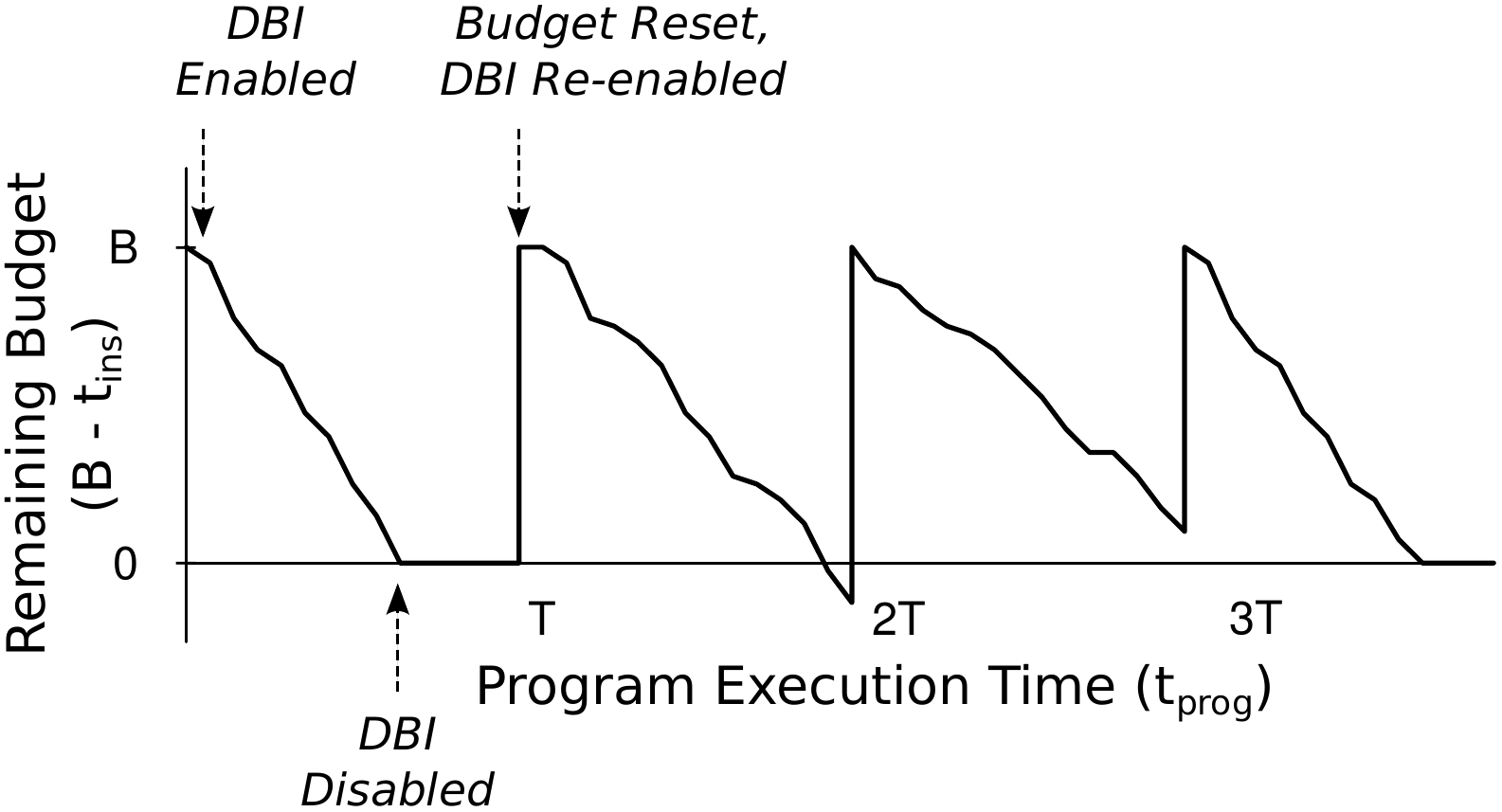}
  \caption{Rate-based DBI~\cite{Arafa2013}.}
  \label{fig:graph}
\end{figure}

Figure~\ref{fig:graph} further illustrates the rate-based DBI approach. 
The X-axis represents the program's execution time $t_{prog}$ and the Y-axis shows the remaining instrumentation budget ($B - t_{ins}$).
The program starts execution in the \enable state i.e., full instrumentation budget is available.
In the first time period $[0,T)$ of the program's execution, instrumentation code executes and reduces the available budget.
Once the instrumentation has fully consumed the budget, the framework will switch to the \disable state and will prevent further instrumentation.
At time $T$, the budget is reset, and the framework returns back to the \enable state~\cite{Arafa2013}.

In~\cite{Arafa2013}, the authors studied three implementations of \oldtool; \tv, \tvc, and \cond.
\tv checks for budget availability at each instrumentation point (through an extra analysis routine) and makes use of Pin's \emph{trace versioning} APIs to enable and disable instrumentation. The implementation of \tv will be discussed later in this section.
\tvc has a similar implementation to \tv but with a reduced frequency of budget checking in the \disable state.
In \cond, the instrumentation routine checks the available budget using a simple \code{if} statement in the instrumentation routine.
Although \tv has the highest budget-checking overhead (compared to the other implementations), it strictly respects the budget and has full budget utilization.
\tvc, compared to \tv, has lower budget-checking overhead but lower budget utilization.
\cond has the lowest budget-checking overhead which leads to loose budget respect.
Also, \cond has another source of high run-time overhead which is the overshoots.
An overshoot occurs when the instrumentation time exceeds the instrumentation budget.
\oldtool shows an average reduction in overhead by 12, 7, and 3  folds, for the three implementations consecutively, compared to native Pin.
Moreover, \tv provides good instrumentation coverage when the instrumentation-budget value is reasonable (10\% for example).
Instrumentation coverage is  the ratio of the output of \oldtool to that of native Pin.
For the reasons mentioned, we focus on \tv in this paper although the suggested approaches work for the other two implementations.

\tv uses Pin's trace versioning APIs to check for budget at each instrumentation point, and enable/disable instrumentation accordingly.
These APIs allow dynamic switching between multiple types (versions) of instrumentation at runtime. 
There is two instrumentation versions in \tool; \code{\vins} refers to enabled instrumentation, and \code{\vbase} when instrumentation is disabled.
When Pin switches versions, it creates a new trace starting from the current instruction.
Listing~\ref{lst:tv} shows a pseudocode outline of \tv (favoring readability over optimality). 

\begin{minipage}[htb]{.9\linewidth}
\lstset{basicstyle=\small, numbers=left, numberstyle=\tiny, firstnumber=1, stepnumber=2, numbersep=2pt}
\begin {lstlisting} [language=c, breaklines=true, frame=lines, label={lst:tv}, caption={Instrumentation rtn of \tv~\cite{Arafa2013}.},captionpos=b,escapeinside={@}{@}]
  For each instrumentation point{
    InsertCall(budget_check);
    if(version == V_BASE) {  
      //check switching to V_INSTRUMENT
      InsertVersionCase(1,V_INSTRUMENT);}
    else if(version == V_INSTRUMENT){ 
      //check switching to V_BASE
      InsertVersionCase(0,V_BASE);}
    switch(version) {
      case V_BASE:	  	
      break; //Do Nothing
    case V_INSTRUMENT:		 
      ...
      InsertCall(analysis_routine);
      break;}
}
\end{lstlisting}
\end{minipage}

To clarify, let \code{Trace\_1} be the sequence of instructions in Listing~\ref{lst:asm1}. 
Assume that \code{Trace\_1} has \code{version = \vins,} and \mbox{\code{``For each instrumentation point''}} means \mbox{\code{``For each instruction''}} in this example.
Pin calls the instrumentation routine at every trace.
The instrumentation routine inserts an inlined call to \mbox{\code{budget\_check()}} before every instruction in the trace.
According to the switch case in Listing~\ref{lst:tv}, the instrumentation routine inserts a call to the analysis routine before every instruction in the trace.
The API \code{InsertVersionCase()} guarantees that the execution of the inserted analysis routine will occur, only if the output of  \mbox{\code{budget\_check()}} matches the ID of the current version (which is $1$ in this case).
Directly after the execution of the instrumentation routine, \mbox{\code{budget\_check()}}, that is inserted before the first instruction, is executed.
Assume that its output is $1$ which means that the budget is currently larger than zero.
In this case, Pin will execute the analysis routine that is inserted before the first instruction.
Afterwards, Pin executes \mbox{\code{budget\_check()}} that is inserted before the second instruction.
Assume that the budget is now fully consumed, so \mbox{\code{budget\_check()}} returns $0$.
Since the output mismatches the ID of the current trace, Pin will switch version to \vbase (i.e. disable instrumentation).
Accordingly, Pin will create a new trace \code{Trace\_2}, with \mbox{\code{version = \vbase}}, starting from the second instruction.
Pin will then execute the instrumentation routine to instrument \code{Trace\_2} according to its version.
Also, the analysis routine inserted before the second and the third instruction will be ignored i.e., not executed.

\begin{minipage}[htb]{.9\linewidth}
\lstset{basicstyle=\small, numbers=none, numberstyle=\tiny, firstnumber=34206, stepnumber=1, numbersep=2pt}
\begin {lstlisting} [language=c, breaklines=true, frame=lines, label={lst:asm1}, caption={Example (1) of a trace.},captionpos=b,escapeinside={@}{@}]
mov eax, dword ptr [rsp+0x30]
and eax, 0x10000000
mov dword ptr [rsp+0x3c], eax 
\end{lstlisting}
\end{minipage}
\section{Redundancy Suppression in \tool}
\label{sec:redun}

Both native Pin and \oldtool generate tracing information that might contain redundancies.
Naik et al. in~\cite{Naik2012} pointed out that, in many applications, output traces contain many redundancies.
For example, in many analysis tools, instrumenting each instruction once is sufficient since the information extracted is the same regardless of the number of times the instruction is executed or instrumented. 
In other words, an instruction can be executed and, hence, instrumented several times. Each time, the same information is extracted causing redundancies.
Examples of these tools include branch profiling tools used for extracting code coverage and memory profiling tools used for building memory access patterns.

Since \oldtool extracts partial tracing information, it can obtain higher instrumentation coverage through the avoidance of tracing redundant information.
To respect the timing properties of a program, \oldtool disables the instrumentation when the instrumentation budget is consumed.
Accordingly, \oldtool generates partial tracing information compared to native Pin.
In other words, there exists a trade-off between the instrumentation budget and the instrumentation coverage.
Hence, multiple runs of \oldtool are required to increase the instrumentation coverage and optimally achieve full coverage.
From a performance point of view, it is preferable to minimize the number of required runs.
For \oldtool, this implies obtaining the maximum possible coverage from each single run without violating the timing constraints.
Accordingly, \oldtool should avoid tracing redundant instrumentation.
In this paper, we specifically focus on the type of analysis tools that do not require tracing redundant information.
\tool should utilize the available instrumentation budget for extracting unique (non-redundant) information.

To prohibit redundant instrumentation, \tool should be able to identify the instrumented code regions. 
The minimum piece of information needed to identify a code region is the starting address.
Thus, the basic idea is to enable \tool to save the starting addresses of instrumented code regions in a log, and \tool should then check the log before instrumenting a new code region.
For the approach to be efficient, \tool should:
\vspace{-\topsep}
\begin{itemize}
 \setlength{\parskip}{0pt}
 \setlength{\itemsep}{0pt plus 1pt}
 \item Prevent re-instrumentation of a code region in the current run and all subsequent runs of the program under analysis.
 \item Avoid increasing run-time overhead.
 \item Avoid creating large-sized logs which increase \tool's memory consumption. Searching a large log may also result in increased run-time overhead.
\end{itemize}
\vspace{-\topsep}
Both steps, saving to the log and searching it, take place in the instrumentation routine, so its overhead is expected to be negligible. 
We avoided adding these steps to the analysis routine which is the main source of overhead in Pin.

In what follows, we discuss our approach for suppressing redundancies in \tool.

\subsection{Granularity of Logged Code Regions}

The first design aspect that we discuss is the granularity of the code regions to be recorded in the log.
We can log addresses of code regions either at the instruction level or the trace level.
It is inefficient to log the address of each instrumented instruction, since
\vspace{-\topsep}
\begin{enumerate}
 \setlength{\parskip}{0pt}
 \setlength{\itemsep}{0pt plus 1pt}
 \item This requires frequent access to the log which adds to run-time overhead.
 \item This results in a large log size which consumes memory.
 \item This leads to searching a large log which can delay program execution and add to the run-time overhead.
\end{enumerate}
\vspace{-\topsep}
An alternative to logging instruction address is to log addresses at a coarser granularity, the trace level.
The instrumentation routine analyzes traces to insert analysis-routine calls.
Recall that a trace is a sequence of program instructions that has a single entry point and may have multiple exit points. 
If Pin detects a jump to an instruction in the middle of a trace, Pin will create a new trace beginning at the target instruction.
So, the instructions inside a trace are always in series i.e., uninterrupted by instructions from another trace.
Thus, \tool will save the trace starting address in addition to the length of the instrumented portion in the trace ($\tuple{\code{Trace Address, Trace Length}}$).
Specifically, \tool will save the relative starting address of the trace with respect to the trace's image.
This guarantees that saved addresses are deterministic between successive runs (especially for the traces of shared libraries).
On the other side, as mentioned earlier in Section~\ref{sec:dime}, trace version switching can cause Pin to create a new trace.
Thus, some trace addresses might only exist in a subset of the runs.


\subsection{Efficient Log Search}
\label{sec:logsearch}
The second design aspect is saving the trace addresses and the trace lengths in a manner that allows for efficient searching of the log.
In this section, we propose three approaches for saving trace addresses and length.
We compare among them qualitatively and quantitatively deriving unexpected results.
We choose one of these approaches to provide \tool with the capability of suppressing redundant instrumentations.

\subsubsection{Hash-Table Log}
The first approach uses a hash-table as the log for saving instrumented traces.
In this approach, \tool saves the trace address to identify instrumented traces.
Whenever \tool instruments a trace, it adds the trace's address to the hash-table.
Also, before instrumenting any trace, \tool searches for the trace address in the hash-table.
Let $A$ be the current trace and $B$ be a trace in the log $L$;
Then, \tool will only instrument $A, \textit{ iff } (address(A) \neq address(B)) \forall B \in L$.
The advantages of using a hash-table for logging traces are:
\vspace{-\topsep}
\begin{itemize}
 \setlength{\parskip}{0pt}
 \setlength{\itemsep}{0pt plus 1pt}
 \item Fast logging (average case: constant; worst case: linear in the hash-table size).
 \item Fast searching (average case: constant; worst case: linear in the hash-table size).
 \item Low number of false negatives (false negatives will occur if \tool prohibits instrumentation of an uninstrumented trace).
 For instance,  let $A$ be the current trace, where $address(A)$ = $100$ and $length(A)$=$80$.
 If the log contains trace $B$, where $address(B)$ = $100$ and $length(B)$=$20$, \tool will prohibit the instrumentation of $A$. 
 Thus, the instructions in the address range $120$ to $180$ will not be instrumented in any run.
\end{itemize}
\vspace{-\topsep}
The disadvantage of this approach is that:
\vspace{-\topsep}
\begin{itemize}
 \setlength{\parskip}{0pt}
 \setlength{\itemsep}{0pt plus 1pt}
 \item Using a hash-table enables \tool to only compare trace addresses while ignoring the trace length.
 This results in false positives.
 A false positive will occur if \tool allows instrumentation of a previously instrumented trace.
 For example, let $A$ be the current trace, where $address(A)$ = $150$ and $length(A)$=$20$.
 \tool may fail to find $A$ in the log, although the log contains trace $B$ such that $address(B)$ = $100$ and $length(B)$ = $80$.
 This means that trace $A$ is previously instrumented as a part of trace $B$.
 Note that one trace being part of another happens due to the creation of new traces through version switching as explained earlier.
\end{itemize}
\vspace{-\topsep}
Section~\ref{sec:eval} presents experimental results that support the listed advantages and disadvantages for the three approaches.
\subsubsection{BST Log}
The second approach is using a binary search tree (BST) to log the addresses of the instrumented traces along with their lengths.
Being sorted, the BST facilitates jumping to a specific range of addresses.
When \tool instruments a trace, it adds the trace to the log such that the trace address is the key and the trace length is the value.
Before instrumenting a trace, \tool searches the BST using the trace address.
If not found, \tool will jump to the log-entry that has the first smaller trace address compared to the current trace address.
\tool will then decide if the current address lies within the trace of the discovered log-entry.
Let $A$ be the current trace and $B$ be a trace in the log $L$.
\tool will not instrument $A$, if $\exists B \in L$ s.t. $(address(B) \leq address(A) < address(B) + length(B))$.
The advantage of using a BST for logging traces is:
\vspace{-\topsep}
\begin{itemize}
 \setlength{\parskip}{0pt}
 \setlength{\itemsep}{0pt plus 1pt}
 \item Less false positives compared to the hash-table approach due to considering the lengths of the logged traces.
\end{itemize}
\vspace{-\topsep}
The disadvantages of this approach, on the other hand, are:
\begin{itemize}[nolistsep,noitemsep]
 \setlength{\parskip}{0pt}
 \setlength{\itemsep}{0pt plus 1pt}
 \item Slower than the hash-table approach in the average case; the complexity of both saving and searching is $log(N)$.
 \item Relatively high false negatives.
 Consider the following example. 
 Assume the current trace is $A = \tuple{100,200}$ (i.e., \code{address}($A$)=100 and \code{length}($A$)=200), and the log entry $B = \tuple{50,80}$ in the log $L$.
 This approach will prevent instrumenting trace $A$ since its starting address lies within the log entry $B$.
 This, however, will consequently prevent \tool from instrumenting the uninstrumented portion of trace $A$ i.e., from address $130$ to address $300$.
\end{itemize}
Additionally, after the program execution and before saving the log to a file for use in subsequent runs, \tool merges directly consecutive traces leading to a smaller log size.
For example, if the log contains two log entries $\tuple{100,50}$ and $\tuple{150,50}$, \tool will merge them into one log entry $\tuple{100,200}$.
Merging log entries decreases the log size and, therefore, reduces the search time leading to less run-time overhead in subsequent runs of \tool.

\subsubsection{Merger BST}
The third approach utilizes a BST as well, but it addresses the second disadvantage of the previous approach.
Using this approach, \tool will prohibit instrumentation, only if the whole current trace is part of a log entry.
Otherwise, \tool allows instrumentation and merges the current trace with the log entry if needed.
Let $A$ be the current trace and $B$ be a trace in the log $L$.
\tool will not instrument $A$, if $\exists B \in L$ s.t. $(address(B) \leq address(A) < address(B) + length(B) \wedge address(A) + length(A) < address(B) + length(B) )$.
For example, let the current trace be $A = \tuple{100,200}$, and the log entry contains $B = \tuple{50,80}$. 
In this approach, \tool instruments trace $A$.
Afterwards, \tool merges trace $A$ and $B$ into one log entry $\tuple{50,250}$ to avoid redundancies in the log. 
The advantages of this approach are:
\vspace{-\topsep}
\begin{itemize}
 \setlength{\parskip}{0pt}
 \setlength{\itemsep}{0pt plus 1pt}
 \item Less false negatives, compared to the BST approach.
\end{itemize}
\vspace{-\topsep}
The disadvantages are:
\vspace{-\topsep}
\begin{itemize}
 \setlength{\parskip}{0pt}
 \setlength{\itemsep}{0pt plus 1pt}
 \item Slower than the hash-table approach in the average case; the complexity of both saving and searching is $log(N)$.
 \item Higher false positives than the BST approach since it allows re-instrumentation of some portions of a trace.
 As mentioned before, Section~\ref{sec:eval} provides experimentation that discusses these observations.
\end{itemize}
\vspace{-\topsep}
Note that a trace address and length will be saved in the log, only if the trace is actually instrumented i.e., the trace's version is \vins.

\subsubsection{Evaluation of the Different Approaches}
\label{sec:eval}
In this section, we describe our experiments to evaluate the three log-search approaches.
According to the results, we choose which approach to extend DIME with the feature of suppressing redundant instrumentation output.

We experiment with two SPEC2006 C benchmark~\cite{spec} programs (\code{lbm} and \code{mcf}) for three runs.
We later, in Section~\ref{sec:exp}, use more SPEC benchmarks to evaluate the performance of modified \tool over up to eight runs.
The experiment uses a branch-profiling analysis tool which has a heavy-weight analysis routine.
It prints out the jump, call, and return instructions in addition to the source address and the destination address.
The tool is based on the $branch\_target\_addr$ pintool that is available as a part of the Pin's kit  v2.12-56759.
The pintool is modified to extract the branch profile of the whole program instead of only a part of it.
A branch profiler is useful for investigating the code coverage of a program.
Each experiment, in this section and Section~\ref{sec:exp}, is conducted once due to the very long execution time when instrumenting the benchmarks on top of native Pin. 
For example, \code{povray} benchmark originally executes in 2.5 minutes, but on top of native Pin, it consumes four days of CPU time. 
The execution time of the other benchmark programs, on top of Native Pin and \tool, will be discussed in Section~\ref{sec:exp}.
However, the runs of each \tool experiment can be considered as repetitions since all the runs operate identically.
In the initialization stage, there exists one minor difference between the first run and the following ones.
The first run starts with an empty log while the following runs read the log from a file before launching and instrumenting the program. 

The evaluation of the proposed approaches is based on the following metrics:
\begin{enumerate}[noitemsep,nolistsep]
 \item \textbf{Instrumentation Coverage}: 
 the ratio of the instrumentation output of \tool to that of native Pin. Note that we consider only unique (non-redundant) traces.
 Increasing the instrumentation coverage is the main objective of the proposed approaches.
 \item \textbf{False positives:} 
 a false positive will occur if \tool permits the instrumentation of a previously instrumented trace or trace-portion.
 This metric measures the ratio of false positives to the total number of instrumented traces in the current run.
 The ratio of false positives indicates the efficiency of the log searching approach in identifying previously instrumented traces.
 As the ratio of false positives decreases, the budget utilization increases and the number of required runs to maintain high coverage decreases.
 \item \textbf{False negatives:} a false negative will take place when \tool refuses to instrument a trace which was not instrumented before.
 The metric measures the ratio of false negatives to the total number of traces that got rejected by \tool in the current run.
 This value includes the trace portions as well i.e., a part of the trace is instrumented but the other part is not.
 As the false negatives increase, the ability of the approach to maintain high coverage decreases.
 \item \textbf{Slow-down factor of the instrumented program}: 
 the ratio of the execution time of the dynamically instrumented benchmark to the execution time of the natively running benchmark.
 This metric examines the ability of \tool to reduce run-time overhead, compared to Pin, while saving to and searching the log to suppress instrumentation redundancy.
 It also checks if the three approaches introduce different run-time overhead.
 Low run-time overhead is essential for the instrumentation of time-sensitive systems as discussed before.
 \item \textbf{Overshoots:} 
 an overshoot will occur when actual instrumentation time exceeds the budget.
 The magnitude of the overshoots shows how strictly modified \tool respects the instrumentation budget.
\end{enumerate}

%
\begin{minipage}[htb]{.9\linewidth}
\lstset{basicstyle=\small, numbers=none, numberstyle=\tiny, firstnumber=34206, stepnumber=1, numbersep=2pt}
\begin {lstlisting} [language=c, breaklines=true, frame=lines, label={lst:asm2}, caption={Example (2) of a trace.},captionpos=b,escapeinside={@}{@}]
test r12, r12
setnz byte ptr [rsp+0x3b]
jnz 0x7ffff7de26a8
mov eax, dword ptr [rsp+0x30]
and eax, 0x10000000
...
call 0x7ffff7df2850
\end{lstlisting}
\end{minipage}
\begin{figure}[htb]
  \centering \includegraphics[width=.8\linewidth]{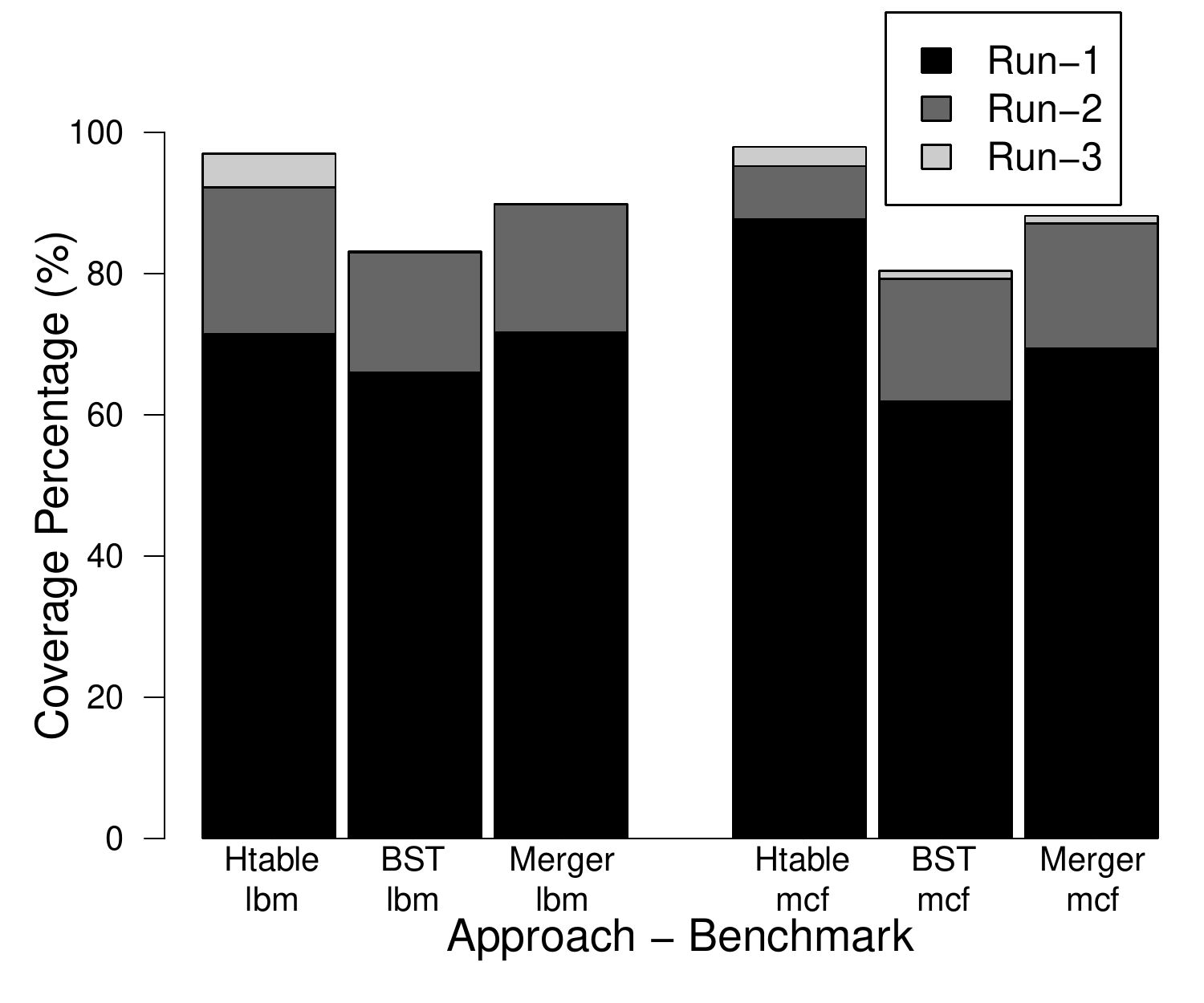}
  \caption{Instrumentation Coverage}
  \label{fig:cov}
\end{figure}

Figure~\ref{fig:cov} shows the instrumentation coverage of the three approaches with \code{lbm} and \code{mcf} consecutively.
The hash-table approach guarantees the highest instrumentation coverage.
After three runs,  it achieves 97\% of the instrumentation coverage of native Pin for \code{lbm} benchmark, and 98\% for \code{mcf}.
The coverage of BST is 83\% and 80\% for \code{lbm} and \code{mcf} consecutively.
Finally, BST-Merger generates 90\% and 88\% of the instrumentation output for \code{lbm} and \code{mcf} consecutively. 
The low ratio of false negatives of the hash-table approach is one reason for achieving the highest coverage.
The hash-table approach is a conservative one which favors re-instrumenting some trace portions over uninstrumenting them.
Also, some scenarios lead to a decreased instrumentation coverage for the BST and Merger BST approaches compared to the hash-table approach.
As mentioned previously, a trace can have multiple exits e.g. can include multiple jump instructions.
Listing~\ref{lst:asm2} is an example of a trace with multiple exits (contains \code{jnz} and \code{call} instructions).
Assume the starting address of the trace is $34192$ and the trace length is $62$.
Assume \tool encounters this trace for the first time, and cannot find the address $34192$ in the log as a key or as a part of another log-entry.
Hence, \tool allows instrumentation of this trace.
Assume that enough instrumentation budget is available to instrument all the instructions in the trace. 
Thus, the trace address along with its length are saved in the log as $\tuple{34192, 62}$.
The instrumentation-routine inserts analysis-routine calls for all the instructions in the trace.
Assume that in the first run of \tool, the first three instructions only execute and a jump (through \code{jnz}) occurs.
In the second run, \tool (BST and BST-Merger) prohibits instrumentation for the trace $34206$  (starting from the $mov$ instruction) since it lies inside the logged trace $\tuple{34192,62}$.
Accordingly, no information is extracted starting from the address $34206$ since these instructions do not execute in the first run and \tool prevents their instrumentation in the following runs.
Although, the program runs with the same inputs, this can occur due to non-deterministic execution of some shared libraries such as libc and the Linux loader. 
For such shared libraries, execution can slightly change according to the processor state.
In such cases, the BST and the Merger BST approaches fail to extract some information, thus decreasing their instrumentation coverage.

\begin{figure}[htb]
	\centering
	\subfloat[False Positives]{\label{fig:pos}\includegraphics[width=0.8\linewidth]{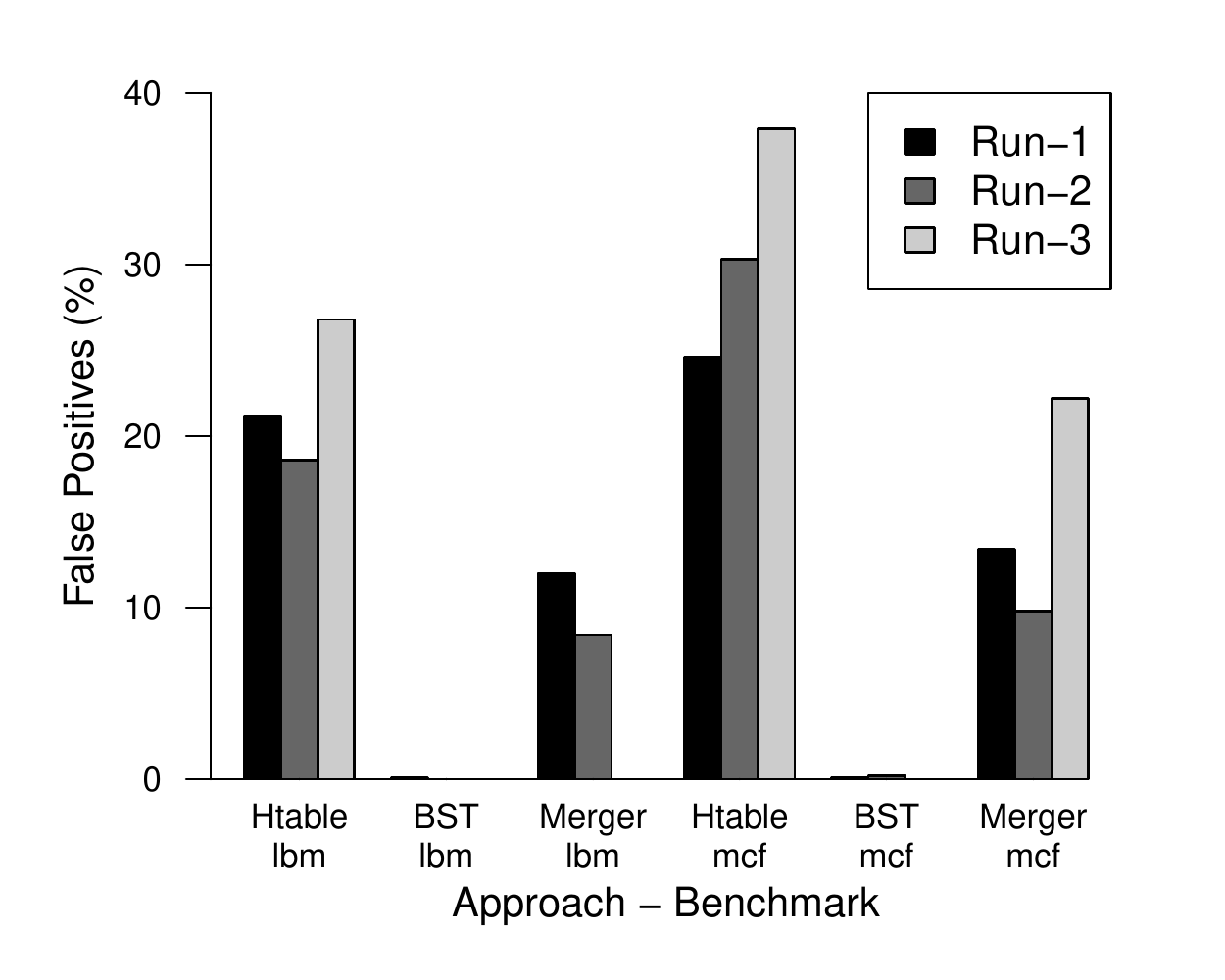}}
	\vskip -4pt
	\subfloat[False Negatives]{\label{fig:neg}\includegraphics[width=0.8\linewidth]{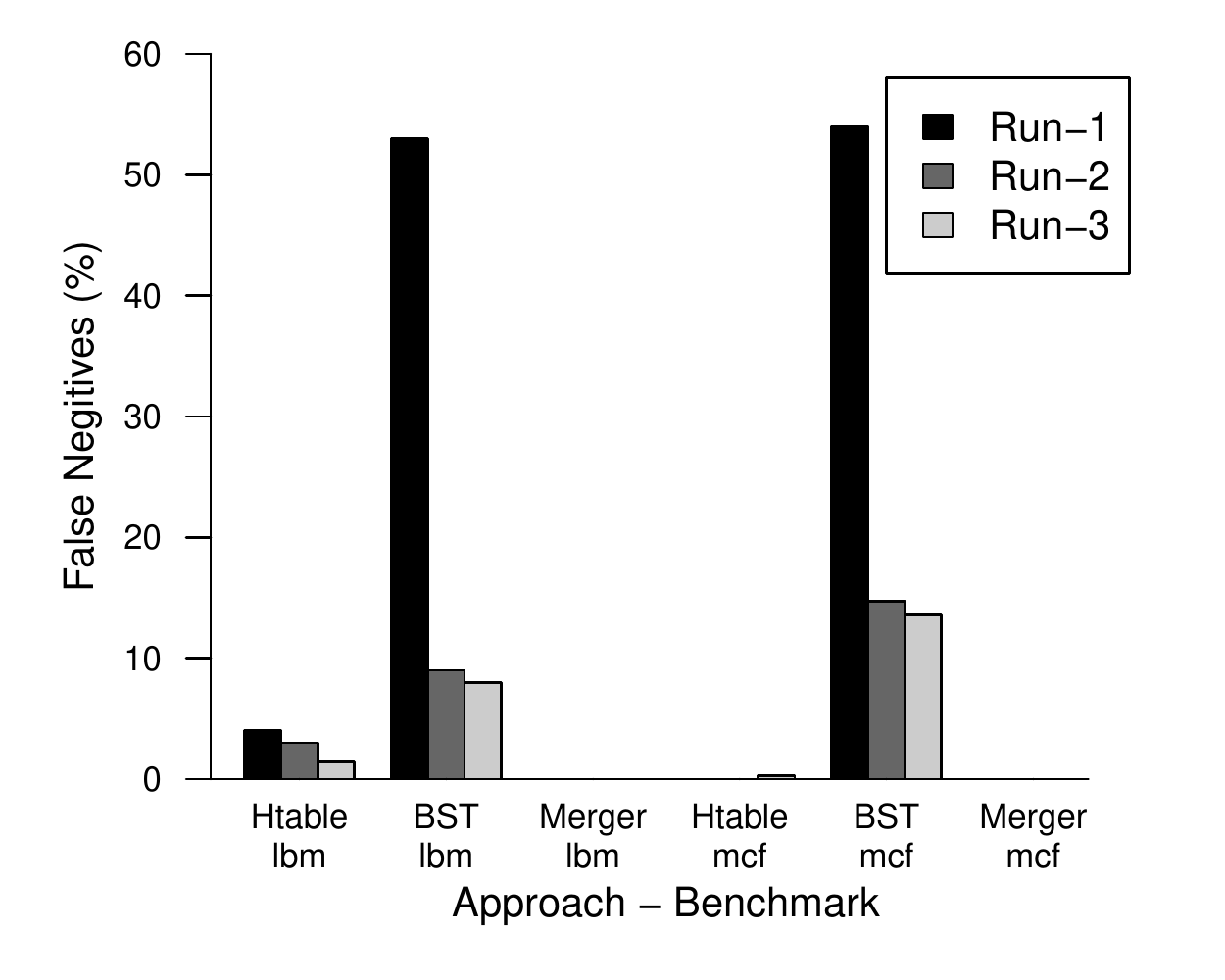}}
 	\caption{Ratio of false positives and ratio of false negatives}
	\label{fig:posneg}
\end{figure}

The ratio of false positives is shown in Figure~\ref{fig:pos}.
The hash-table approach has the highest ratio with both \code{lbm} and \code{mcf} benchmarks.
The BST-Merger approach has moderate values of false positives, whereas BST has approximately zero false positives.
This means that BST accurately identifies the previously instrumented traces and efficiently utilizes the budget to instrument other traces.
On the other hand, BST has a high ratio of false negatives, as shown in Figure~\ref{fig:neg}, which is an undesirable feature.
BST-Merger sustains approximately zero false negatives, and hash-table has negligible values of false negatives ratio.
The scenarios discussed in Section~\ref{sec:logsearch} explain the values in Figures~\ref{fig:pos} and~\ref{fig:neg}.
Note that false-negatives ratio is more critical than false positives.
Although false positives cause instrumentation redundancies, it is safer.
False negatives prevent code portions from being instrumented in any run which can dramatically decrease the instrumentation coverage.
The results show that there is a trade-off between false positives and false negatives.
In such case, we prefer the approach that maintains low ratio of false negatives even if it has high ratio of false positives.
Thus, the hash-table and the BST-Merger approaches outperform the BST one.

\begin{figure}[htb]
	\centering
	\subfloat[lbm benchmark]{\label{fig:lbmx}\includegraphics[width=0.8\linewidth]{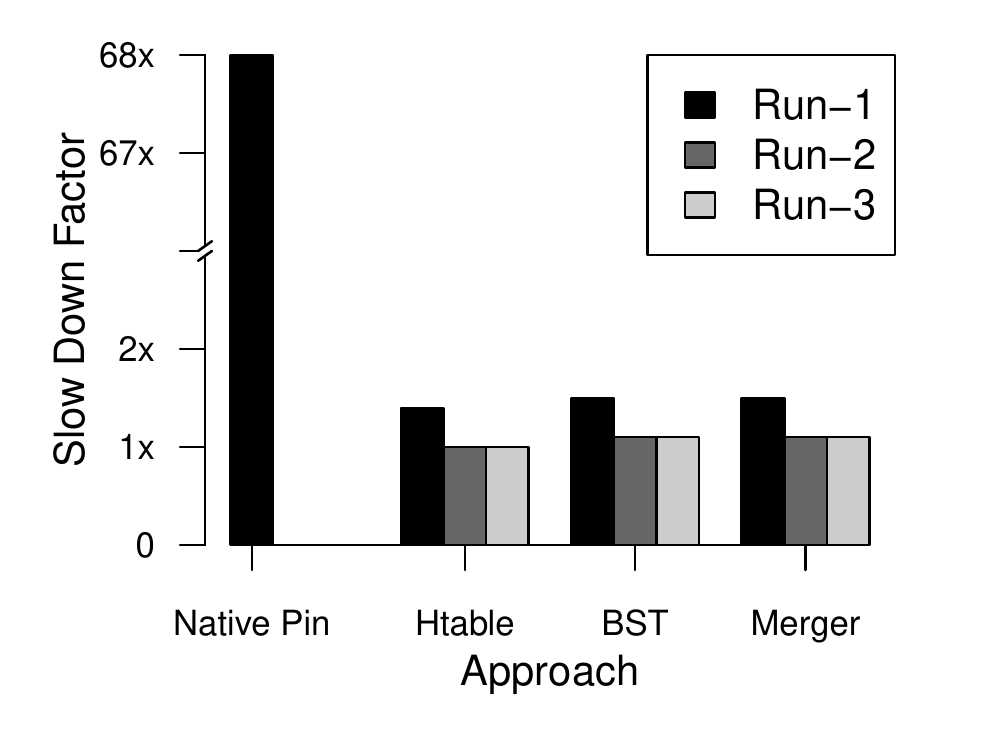}}
	\vskip -3pt
	\subfloat[mcf benchmark]{\label{fig:mcfx}\includegraphics[width=0.8\linewidth]{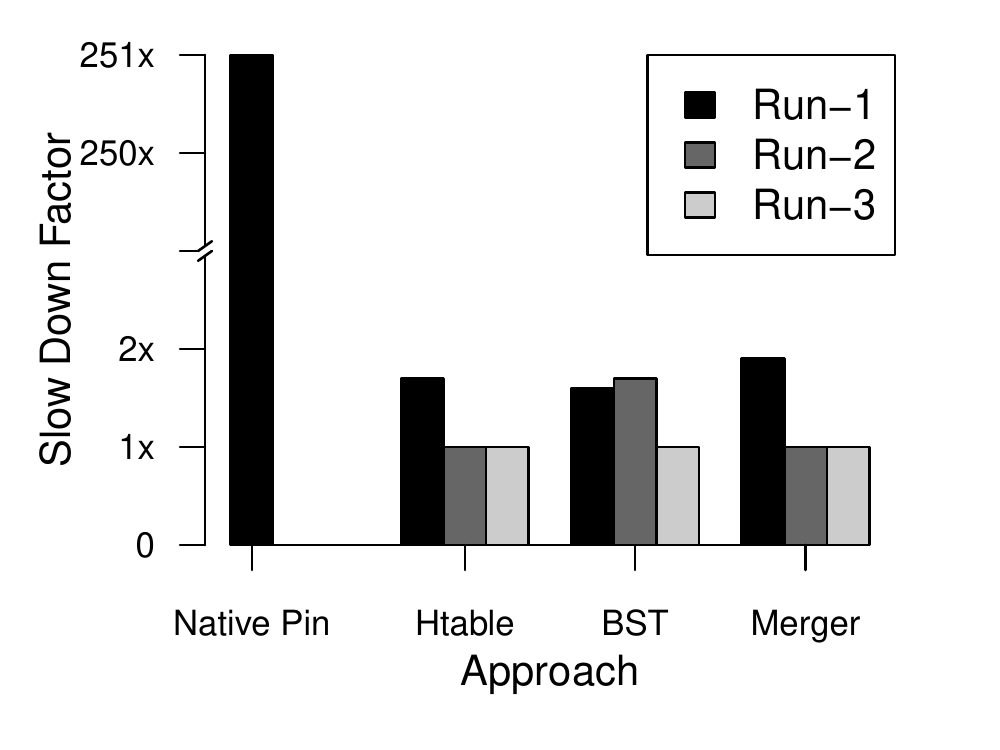}}
	\caption{Slow-down factors.}
	\label{fig:slowdown}
\end{figure}

Figure~\ref{fig:lbmx} presents the slow-down factors of native Pin and the proposed approaches of \tool with \code{lbm} benchmark.
On top of native Pin, \code{lbm} runs 68x slower than the native execution.
On the other hand, the hash-table approach of Pin achieves a slow-down of 1.4x, 1x, and 1x for three consecutive runs.
The overhead of the BST approach is 1.5x, 1x, and 1x, while that of BST-Merger is 1.5x, 1x, 1x for three runs.
In Figure~\ref{fig:mcfx}, native Pin slows down the execution 251x with \code{mcf} benchmark.
Whereas, the slow-down factors of the hash-table approach for the three runs are 1.7x, 1x, and 1x.
These of the BST are 1.6x, 1.7x, and 1x, and BST-Merger achieves slow-down of 1.9x, 1x, and 1x.
To sum up, \tool reduces the run-time overhead by at least 45 folds for \code{lbm} and 132 folds for \code{mcf}.
These numbers reveal that the three modifications of \tool are able to dramatically reduce the run-time overhead of native Pin.
Thus, all of the three \tool modifications are suitable for dynamically instrumenting time-sensitive systems.
Comparing the three approaches to each other, none of them shows a significant overhead-decrease over the others.
Consequently, run-time overhead is not a factor that differentiates among the three approaches.

\begin{figure*}[htb]
	\centering
	\subfloat[Hash-table]{\label{fig:oversht_htable}\includegraphics[width=0.3\linewidth]{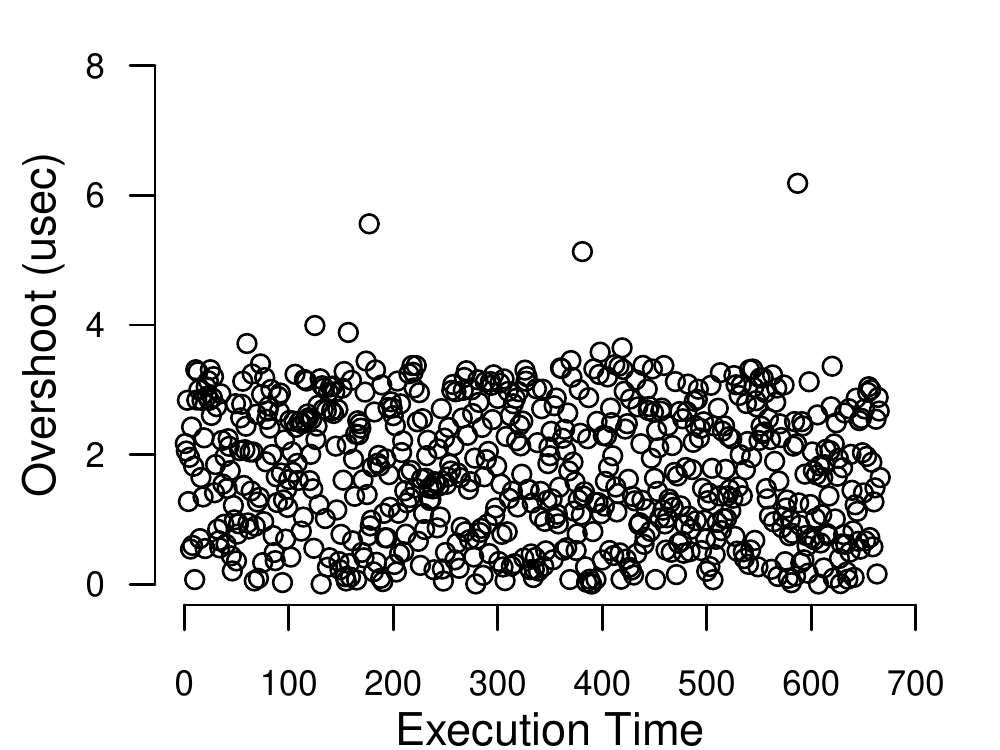}}
	\subfloat[BST]{\label{fig:oversht_bstree}\includegraphics[width=0.3\linewidth]{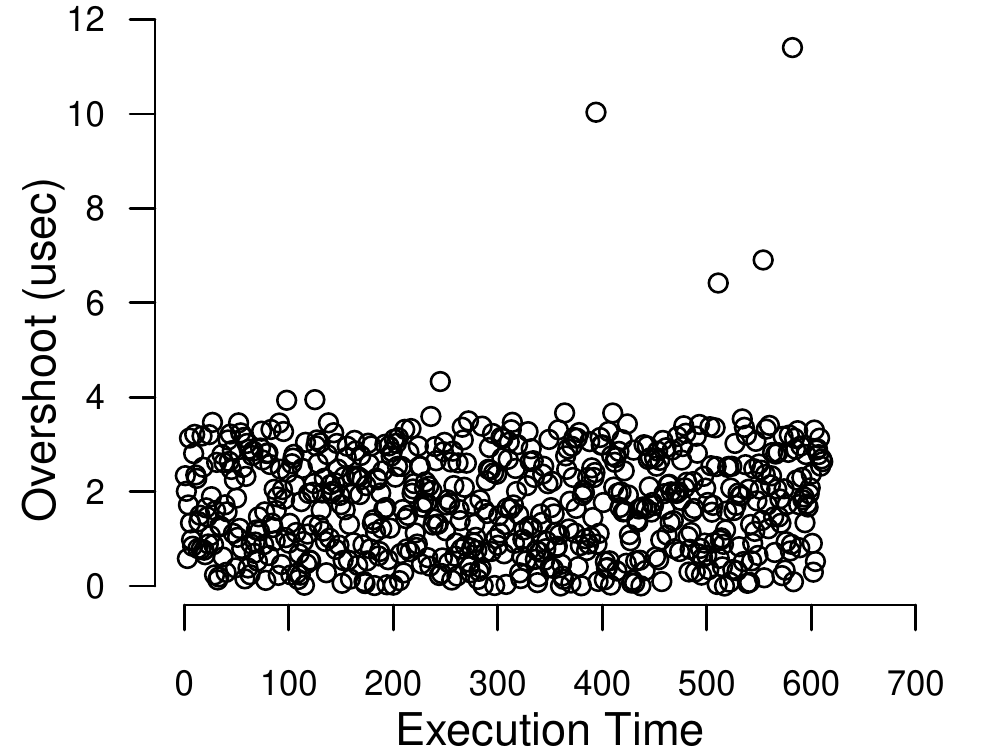}}
	\subfloat[Merger BST]{\label{fig:oversht_optimal}\includegraphics[width=0.3\linewidth]{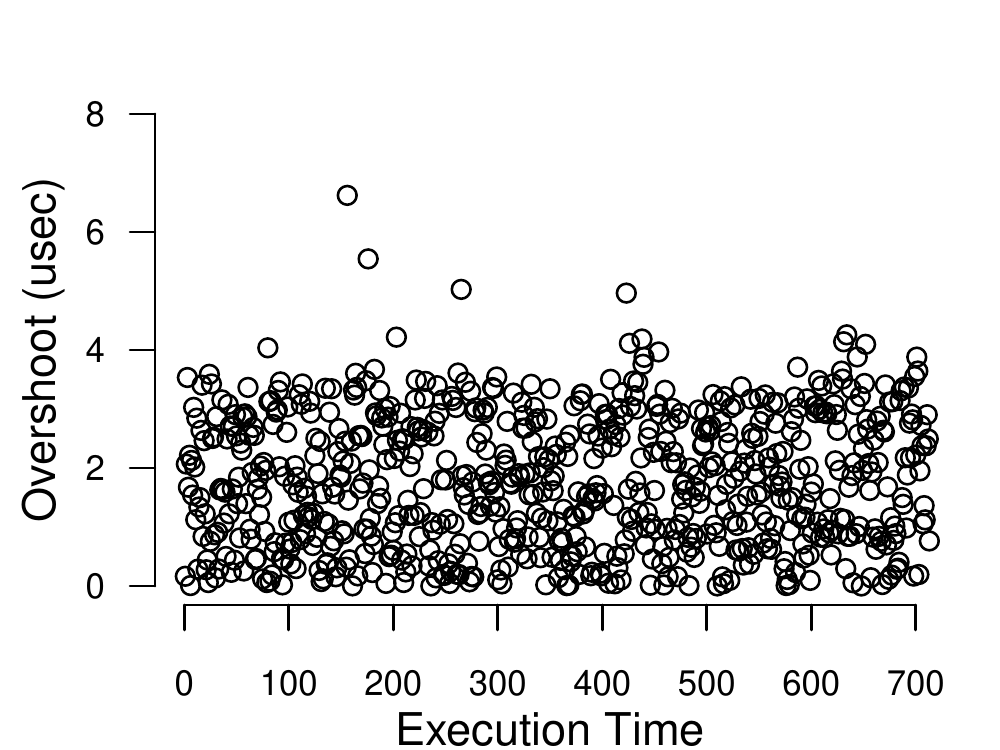}}
	\caption{Overshoots of the three approaches in the first run with the \code{mcf} benchmark.}
	\label{fig:oversh}
\end{figure*}

Figure~\ref{fig:oversh} shows the overshoots' magnitude for the three proposed approaches of \tool over the execution time of the \code{mcf} benchmark while instrumenting it using the \tool version of the branch-profiling pintool.
The three approaches respect the instrumentation budget; the values for the most frequent overshoots lie below 4 microseconds.
There is not significant differences in the overshoots' magnitudes among the three approaches of \tool.
Thus, this metric is also not a factor to favor one approach over the others.


Non-intuitively, the evaluation metrics reveal that the simplest approach, which is the hash-table one, results in the best instrumentation coverage results.
Moreover, the hash-table approach provides low values of false negatives, maintains low run-time overhead, and respects the instrumentation budget.
Accordingly, we choose the hash-table approach to support instrumentation-redundancy suppression in \tool.

\section{Performance Evaluation}
\label{sec:exp}
This section presents the experimentation of \tool and discusses its performance.
\subsection{Experimental Setup}
We experiment with the SPEC2006 benchmark suite~\cite{spec} including C and C++ integer and floating point programs.
The experiments run on top of a workstation hosting a quad-core i7 3.4 GHz Intel processors with 8 MB of cache, and 16 GB of RAM.
The operating system is an Ubuntu 12.04 patched with a real-time kernel v3.2.0-23 to convert Linux into a fully preemptible kernel.
To obtain accurate results, we inhibit task migration between cores and lock core speed to the maximum frequency.
The experimentation environment also maintains a real-time scheduling policy and priority.
These modifications guarantee accurate results for performance evaluation and are not mandatory for \tool correctness.
The instrumentation objective is extracting the branch-profile of the program.
We use the Pin kit v2.12-56759 and gcc v4.6.3.
The time-period parameter is set to one second and the instrumentation budget is set to 0.1 seconds for all the benchmark programs.
Note that the experimentation included more benchmarks, however, these extra benchmarks are not reported since the execution time on top of native Pin exceeded twenty days.

We evaluate the performance of \tool using the following metrics:
\vspace{-\topsep}
\begin{enumerate}
 \setlength{\parskip}{0pt}
 \setlength{\itemsep}{0pt plus 1pt}
 \item \textbf{Instrumentation Coverage:} the ratio of the amount of unique extracted traces by \tool to the amount of those extracted by native Pin.
 This metric demonstrates the ability of \tool to extract the maximum possible amount of information while respecting the instrumentation budget.
 \item \textbf{Slow-down factor of the instrumented program:} the ratio of the execution time of the dynamically instrumented benchmark to its original execution time (without instrumentation).
 This metric reflects the reduction of run-time overhead of \tool compared to native Pin.
\end{enumerate}
\vspace{-\topsep}

\subsection{Experimental Results}

\begin{figure}[htb]
  \centering \includegraphics[width=.9\linewidth]{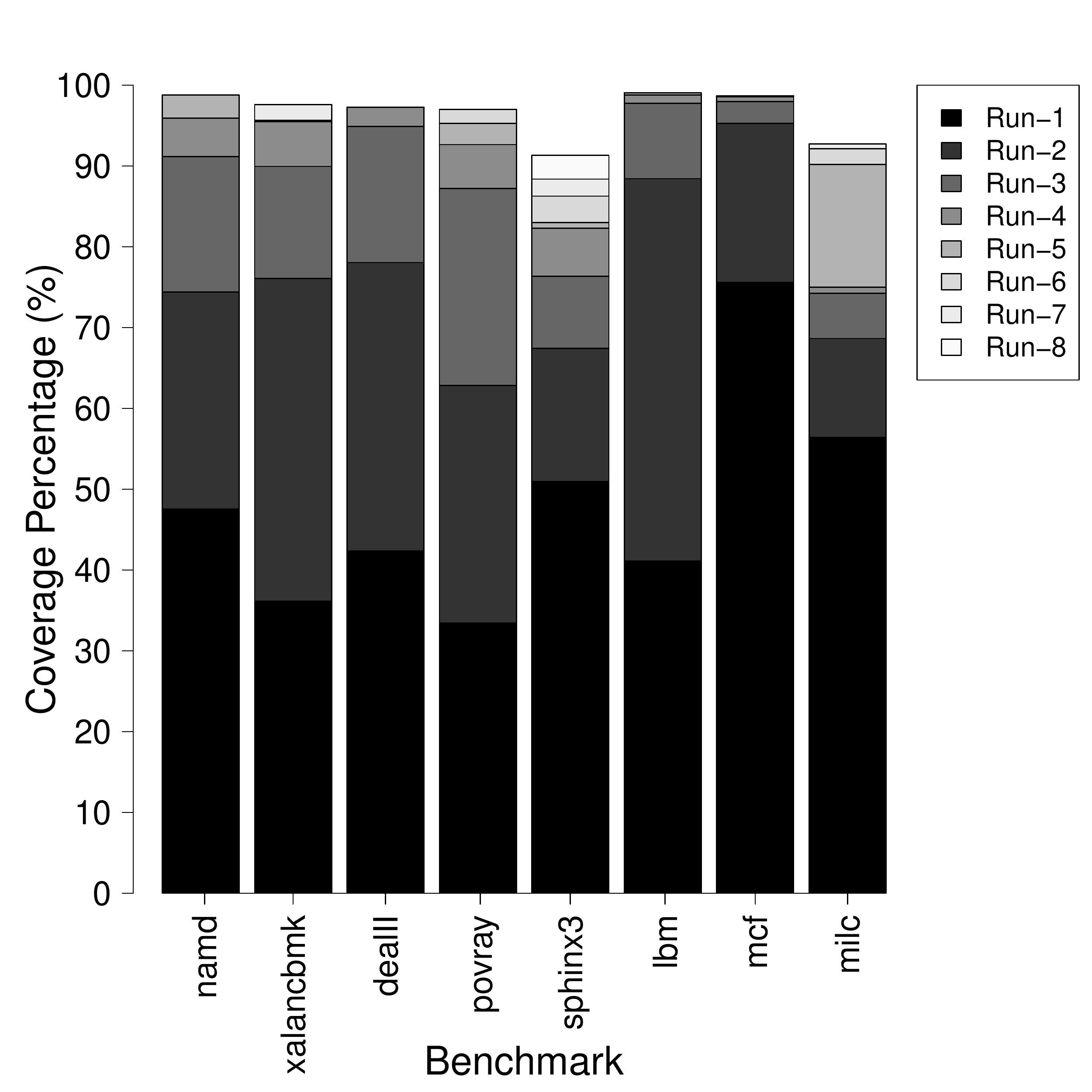}
  \caption{Instrumentation Coverage}
  \label{fig:covall}
\end{figure}
\tool is capable of maintaining high instrumentation coverage.
Figure~\ref{fig:covall} shows the ratio of the amount of the extracted instrumentation output through multiple runs of \tool with respect to that extracted by native Pin.  
\tool is capable of extracting 97\% and 99\% of the instrumentation output in four runs for \code{dealII} and \code{mcf} benchmarks consequently.
In five runs, \tool generates a coverage of 99\% for both \code{namd} and \code{lbm}.
It extracts 92\% and 97\% in the sixth run for \code{milc} and \code{povray} consequently.
\tool also extracts 98\% when instrumenting \code{xalancbmk} for seven runs, and 91\% of the instrumentation output of \code{sphinx3} after eight runs.

\begin{figure}[htb]
  \centering \includegraphics[width=\linewidth]{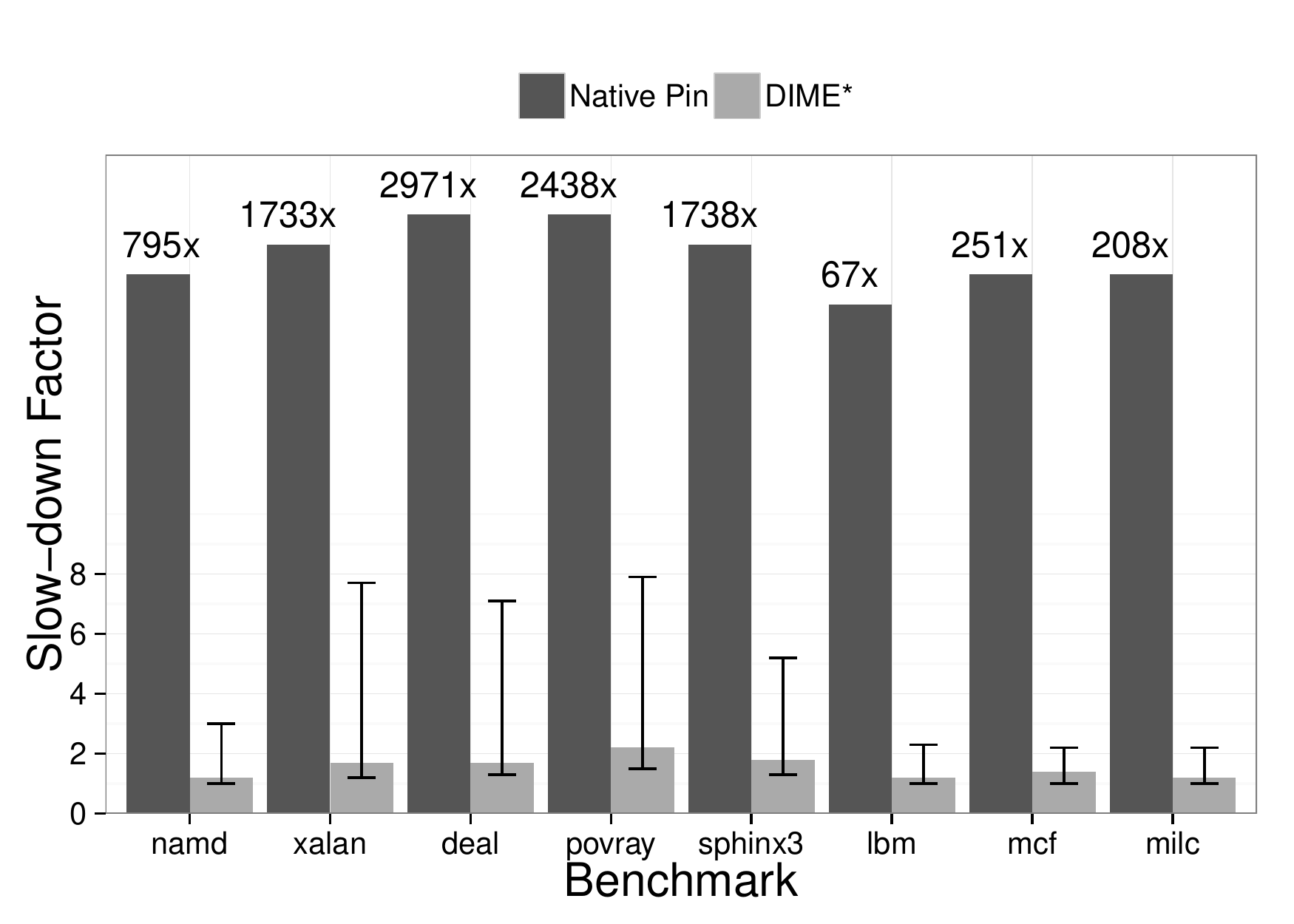}
  \caption{Slow-down factors of native Pin and \tool.}
  \label{fig:execall}
\end{figure}
\tool outperforms native Pin in terms of run-time overhead.
It reduces the run-time overhead of dynamic instrumentation by one to three orders of magnitude compared to native Pin.
Figure~\ref{fig:execall} shows the slow-down factor of native Pin, and the average slow-down factor of the eight runs of \tool for each benchmark program.
Note that the average slow-down factor is the geometric mean.
Native Pin dramatically slows down the program execution.
The average slow down of native Pin is 706x the original benchmark execution, with maximum value of 2971x and minimum value of 67x.
The benchmarks' continuous execution time on top of native Pin ranges from four hours to nine days with average of four days.
On the other hand, the benchmarks on top of \tool takes from three to 42 minutes with average of nine minutes.
The average slow-down of \tool is 1.5x, with maximum value of 8x and minimum value of 1x.

Running a program multiple times on top of \tool is less time-consuming than native Pin.
More importantly, \tool allows the program to maintain its timing properties (or extra-functional properties in general). 
Respecting such properties is essential for time-sensitive systems.
Additionally, multiple runs of \tool provide very high instrumentation coverage compared to native Pin.

%
\section{Case Studies}
\label{sec:case}
This section demonstrates the scalability and the practicality of \tool through two case studies.

\subsection{VLC Case Study}
This case study reveals the usability of \tool for instrumenting multi-threaded time-sensitive software.
VLC~\cite{VLC}, developed by the VideoLan organization, is a multi-platform media player.
It consists of approximately \num{600000} lines of code and uses dependencies of approximately three million lines of code.
In this case study, we aim to extract the call context tree of VLC  v2.0.8 while playing  a high definition, 29.97 fps, 720x480, 1 Mbps bitrate video.
The call context tree is useful for performance analysis and applying program optimizations~\cite{Serrano}.
The tool, used in this case study, is based on the DebugTrace pintool which is available in Pin's v2.12 kit.
We use the tool to extract the call traces, then we use the call traces to build a call context tree.
The platform is an Ubuntu 12.04 machine hosting a quad-core i7 2.6 GHz Intel processors with 8 GB of RAM.
For \tool, we set the time period to one second and the budget to 10\% which enables VLC to run the video smoothly.

Table~\ref{tab:vlc} shows the number of video blocks that VLC decodes for viewing frames, and the extracted percentage of the nodes and the edges of the call-context tree. 
The table lists the number of decoded blocks without instrumentation, with native Pin, and with \tool.
It also shows the coverage of \tool compared to Pin.

Our conclusion is that \tool dramatically decreases the run-time overhead of dynamic instrumentation and maintains almost the same instrumentation coverage.
On top of native Pin, VLC fails to maintain a continuous video playback. 
The video is unwatchable; only 37 video blocks (out of 594) are decoded.
VLC generates multiple errors of dropping video frames due to very slow processing.
On the other hand, \tool enables VLC to play the video continuously.
VLC decodes 574, 594, and 594 video blocks (out of 594) for the three runs consecutively.
\tool extracts 98\% of the call context tree nodes and 96\% of the edges.

\begin{table}[htb]
 \begin{tabular}{l  c  c c}
				      & Decoded Blocks & Nodes(\%) & Edges(\%) \\ \hline \hline
 Original 				&  	594		& N/A & N/A \\ \hline
 Native Pin 			&   	37		&  100\% &  100\% \\ \hline
 \tool (run 1)	  		 & 	574		& 95.7\% & 93.6\%\\ 
\hspace{10.5 mm}(run 2)	  &  	594		& 98\% 	& 96.2\%\\
\hspace{10.5 mm}(run 3)	  & 	594		& 98.2\% & 96.5\%\\ \hline
 \end{tabular}
 \caption{Results for the VLC case study}
   \label{tab:vlc}
\end{table}


\subsection{PostgreSQL Case Study}
\tool is useful for extracting sufficient analysis information while maintaining high quality of service (QoS).
PostgreSQL~\cite{postgres} is a powerful object-relational database management system that supports SQL standards.
PostgreSQL $9.1$ consists of approximately one million lines of code.
When an application sends a database request to PostgreSQL, a connection will be established, and a parser will check the query syntax and create a query tree.
Then, an optimizer generates a query plan from the tree and sends the plan to the executor.
Finally, the executor steps through the commands in the query plan to retrieve the required information or to make the required updates to the database.
Pgbench~\cite{pgbench} is a benchmarking tool for testing the performance of PostgreSQL.
It runs a sequence of transactions multiple times in multiple database sessions.
The objective of this case study is the extraction of the branch-profile of PostgreSQL while processing a total of \num{800000} transactions.
PostgreSQL runs 16 database sessions where each session consists of \num{50000} transactions.
The case study runs on an Ubuntu 12.04 platform hosting a quad-core i7 2.6 GHz Intel processors with 8 GB of RAM.
We set the budget of \tool to 7\% and the time period to one second. These values enable \tool to extract sufficient information while keeping high performance of PostgreSQL.

Table~\ref{tab:postgres} shows the performance of PostgreSQL while (1) running natively, (2) running on top of native Pin, and (3) running on top of \tool.
The total time consumed to process the \num{800000} transactions is an indication of the performance.
As the processing time increases, the degradation in the quality of service (QoS) increases. 
The processing time reported excludes connection-establishing time.
The table also shows the coverage of \tool with respect to that of Pin.

In this case study, \tool extracts  97\% of the analysis data while reducing the run-time overhead by 68 folds compared to native Pin.
Originally, PostgreSQL processes all the transactions in 46 seconds.
Native Pin causes a slowdown of 96x; PostgreSQL executes the same number of transactions in 1.2 hours.
On the other hand, \tool maintains slowdown of only 1.4x, 1.4x, and 1.3x in three runs consecutively.

\begin{table}[htb]
 \begin{tabular}{l  c  c}
				& Total Time (sec) & Coverage(\%) \\ \hline \hline
 Original 			& 46 			& N/A  \\ \hline
 Native Pin 		& \num{4419}  &  100\%\\ \hline
 \tool (run 1)	  	 & 65 			& 42\% \\ 
\hspace{10.5 mm}(run 2)	  & 65 			& 75\% \\
\hspace{10.5 mm}(run 3)	  & 61			& 97\% \\ \hline
 \end{tabular}
 \caption{Results for the PostgreSQL case study}
   \label{tab:postgres}
\end{table}

\section{Related Work}
\label{sec:related}

A program can be instrumented at the source code level either automatically or manually.
In automatic instrumentation, a tool parses the program, may generate a CFG, and eventually insert instrumentation points.
Multiple works~\cite{Fischmeister2010a,Kashif2012,Kashif2013} investigate static source-code time-aware instrumentation tools.
On the other hand, manual instrumentation requires that the developer specifies the instrumentation locations~\cite{Simon2008}.
Manual instrumentation is highly flexible, but the induced effect of instrumentation on the timing behavior is hard to estimate by the developer.

Some instrumentation tools are also capable of inserting instrumentation points to binary executables, either statically or dynamically.
QPT~\cite{Larus1993}, EEL~\cite{Larus1995}, and ATOM~\cite{Srivastava2004} are examples of static binary instrumentation tools.
Static instrumentation is based on static analysis and, hence, cannot react to application changes at run time.
Dynamic binary instrumentation, on the other hand, does not require any pre-processing of the program under analysis.

Example of dynamic binary instrumentation tools that use code transformation during program execution include Dyninst~\cite{Buck2000} and Vulcan~\cite{Edwards2001}.
Most of these instrumentation tools, however, modify the native behavior of the program under analysis~\cite{Bruening2012}.
Other tools have software code caches and are able to dynamically compile binaries such as Pin~\cite{Luk2005}, DynamoRIO~\cite{Bruening2003}, and Valgrind~\cite{Nethercote:2007}.

Superpin~\cite{Wallace2007} is a parallelized version of Pin that reduces the overhead of dynamic binary instrumentation by utilizing multiple cores.
The performance gain is limited by the number of cores and shared memory structures.
Yu et al.~\cite{Yu2011} introduce MT-profiler, a multi-threaded profiling framework built on top of DynamoRIO to instrument parallel programs.
Zhao et al.~\cite{Zhao2010} introduce PiPA, a pipelined profiling and analysis tool for parallelizing dynamic instrumentation on multi-core platforms.
Moseley et al~\cite{Moseley2007} present a probe-based application monitor which is injected to the original application under analysis.
The approach we present in this work can make use of such parallelization techniques to reduce overhead as well.

Upton et al.~\cite{Upton2009} introduce a buffering system for Pin to reduce overhead.
Kumar et al.~\cite{Kumar2005} present an instrumentation optimizer for program monitoring and profiling.
While these instrumentation approaches focus on reducing the overhead of instrumentation, they are unaware of the program's timing requirements.
They, however, are also orthogonal to our proposed approach.

The work by Arnold and Ryder~\cite{Arnold2001} is the most relevant to reducing the overhead of dynamic instrumentation following \oldtool.
The authors' approach involves duplicating code regions and using counter-based sampling to switch between the instrumented and non-instrumented versions of the code.
Code duplication results in a large increase in code space.
Since event-based sampling only samples events according to their frequency of occurrence, this results in a reduced instrumentation overhead.
However, as explained in~\cite{Navabpour3}, event-based sampling can result in sampling bursts which can cause high degradation in performance
Other sampling-based approaches are also used for performance optimizations~\cite{Froyd2005}.
These approaches either apply optimizations specific to the instrumentation objective or use compiler-specific information to perform optimizations.

\section{Conclusion}
\label{sec:conc}

Dynamic binary instrumentation tools are used to analyze program behavior and extract debugging information at runtime.
Most applications, however, cannot leverage existing tools for analysis or debugging purposes due to the performance degradation resulting from the instrumentation process.
Time-sensitive applications, for instance, require a bound on the execution time overhead.

In this paper, we propose \tool that performs instrumentation only within a user-specified instrumentation budget.
\tool suppresses logging of redundant information to reduce the instrumentation overhead.
The results show a reduction in the overhead of the instrumentation process by one to three orders of magnitude compared to native Pin while achieving up to 99\% of the instrumentation coverage.
\tool was able to extract 97\% of the call context tree of the VLC video player while playing a high definition video.
VLC fails to provide a watchable video while being instrumented using native Pin.
\tool was also used for the branch profiling of the PostgreSQL database management system and was able to extract 97\% of the instrumentation information in three runs.
\tool extracts this information in less than two minutes per run while native Pin takes 1.2 hours to extract the information.
The presented case studies show the scalability of \tool and its ability to limit the instrumentation overhead while achieving a high instrumentation coverage.

\bibliographystyle{abbrv}
\bibliography{refs}

\begin{thebibliography}{10}

\bibitem{pgbench}
{Pgbench: Benchmarking Tool for PostgreSQL}.
\newblock http://wiki.postgresql.org/wiki/Pgbench.

\bibitem{postgres}
{PostgreSQL Global Development Group}.
\newblock http://www.postgresql.org/.

\bibitem{VLC}
{VLC Media Player}.
\newblock http://www.videolan.org/vlc/index.html.

\bibitem{Arafa2013}
P.~Arafa, H.~Kashif, and S.~Fischmeister.
\newblock Dime: Time-aware dynamic binary instrumentation using rate-based
  resource allocation.
\newblock In {\em Proc. of the 13th International Conference on Embedded
  Software (EMSOFT)}, Montreal, Canada, Sept 2013.

\bibitem{Arnold2001}
M.~Arnold and B.~G. Ryder.
\newblock {A Framework for Reducing the Cost of Instrumented Code}.
\newblock In {\em Proc. of the ACM SIGPLAN Conf. on Programming language design
  and implementation (PLDI)}, 2001.

\bibitem{Bock2011}
S.~Bock, B.~Childers, R.~Melhem, D.~Mosse, and Y.~Zhang.
\newblock {Analyzing the Impact of Useless Write-backs on the Endurance and
  Energy Consumption of PCM Main Memory}.
\newblock In {\em Performance Analysis of Systems and Software (ISPASS), 2011
  IEEE International Symposium on}, pages 56--65, 2011.

\bibitem{Navabpour3}
B.~Bonakdarpour, S.~Navabpour, and S.~Fischmeister.
\newblock {Sampling-based Runtime Verification}.
\newblock In {\em Proc. of the 17th Intl. Conf. on Formal Methods (FM)}, Jun.
  2011.

\bibitem{Bruening2003}
D.~Bruening, T.~Garnett, and S.~Amarasinghe.
\newblock {An Infrastructure for Adaptive Dynamic Optimization}.
\newblock In {\em Proc. of the Intl. Symp. on Code Generation and Optimization
  (CGO)}, 2003.

\bibitem{Bruening2012}
D.~Bruening, Q.~Zhao, and S.~Amarasinghe.
\newblock {Transparent Dynamic Instrumentation}.
\newblock {\em SIGPLAN Not.}, 47(7), Mar. 2012.

\bibitem{Buck2000}
B.~Buck and J.~K. Hollingsworth.
\newblock {An API for Runtime Code Patching}.
\newblock {\em Int. J. High Perform. Comput. Appl.}, 14(4), Nov. 2000.

\bibitem{Edwards2001}
A.~Edwards, H.~Vo, and A.~Srivastava.
\newblock {Vulcan: Binary Transformation in a Distributed Environment}.
\newblock Technical report, 2001.

\bibitem{Fischmeister2010a}
S.~Fischmeister and P.~Lam.
\newblock {Time-Aware Instrumentation of Embedded Software}.
\newblock {\em IEEE Transactions on Industrial Informatics}, 2010.

\bibitem{Froyd2005}
N.~Froyd, J.~Mellor-Crummey, and R.~Fowler.
\newblock {Low-overhead Call Path Profiling of Unmodified, Optimized Code}.
\newblock In {\em Proc. of the 19th Annual Intl. Conf. on Supercomputing
  (ICS)}, 2005.

\bibitem{spec}
J.~L. Henning.
\newblock {SPEC CPU2000: Measuring CPU Performance in the New Millennium}.
\newblock {\em Computer}, 33(7), 2000.

\bibitem{Kashif2013}
H.~Kashif, P.~Arafa, and S.~Fischmeister.
\newblock {INSTEP: A Static Instrumentation Framework for Preserving
  Extra-functional Properties}.
\newblock In {\em Proc. of the 19th IEEE Intl. Conf. on Embedded and Real-Time
  Computing Systems and Applications (RTCSA)}, Aug. 2013.

\bibitem{Kashif2012}
H.~Kashif and S.~Fischmeister.
\newblock Program transformation for time-aware instrumentation.
\newblock In {\em Proc. of the 17th IEEE Intl. Conf. on Emerging Technologies
  \& Factory Automation (ETFA)}, Sep. 2012.

\bibitem{Kim2004}
M.~Kim, M.~Viswanathan, S.~Kannan, I.~Lee, and O.~Sokolsky.
\newblock {Java-MaC: A Run-Time Assurance Approach for Java Programs}.
\newblock {\em Form. Methods Syst. Des.}, 2004.

\bibitem{Kumar2005}
N.~Kumar, B.~R. Childers, and M.~L. Soffa.
\newblock {Low Overhead Program Monitoring and Profiling}.
\newblock In {\em Proc. of the 6th ACM SIGPLAN-SIGSOFT workshop on Program
  Analysis for Software Tools and Engineering (PASTE)}, 2005.

\bibitem{Larus1993}
J.~Larus.
\newblock Efficient program tracing.
\newblock {\em Computer}, 26(5), 1993.

\bibitem{Larus1995}
J.~R. Larus and E.~Schnarr.
\newblock {EEL: Machine-Independent Executable Editing}.
\newblock {\em SIGPLAN Not.}, 30, 1995.

\bibitem{Luk2005}
C.-K. Luk, R.~Cohn, R.~Muth, H.~Patil, A.~Klauser, G.~Lowney, S.~Wallace, V.~J.
  Reddi, and K.~Hazelwood.
\newblock {Pin: Building Customized Program Analysis Tools with Dynamic
  Instrumentation}.
\newblock In {\em Proc. of the ACM SIGPLAN Conf. on Programming Language Design
  and Implementation (PLDI)}, 2005.

\bibitem{Mellor-Crummey1989}
J.~M. Mellor-Crummey and T.~J. LeBlanc.
\newblock {A Software Instruction Counter}.
\newblock In {\em Proc. of the 3rd Intl. Conf. on Architectural Support for
  Programming Languages and Operating Systems (ASPLOS)}, 1989.

\bibitem{Moore2003}
L.~J. Moore and A.~R. Moya.
\newblock {Non-Intrusive Debug Technique for Embedded Programming}.
\newblock In {\em Proc. of the 14th Intl. Symp. on Software Reliability
  Engineering (ISSRE)}, 2003.

\bibitem{Mork}
P.~Mork.
\newblock {Techniques for Debugging Parallel Programs}.
\newblock Technical report, University of Miskolc.

\bibitem{Moseley2007}
T.~Moseley, A.~Shye, V.~Reddi, D.~Grunwald, and R.~Peri.
\newblock {Shadow Profiling: Hiding Instrumentation Costs with Parallelism}.
\newblock In {\em Intl. Symp. on Code Generation and Optimization(CGO)}, Mar.
  2007.

\bibitem{Mytkowicz2008}
T.~Mytkowicz, A.~Diwan, M.~Hauswirth, and P.~Sweeney.
\newblock {We have it Easy, but do we have it Right?}
\newblock {\em IEEE Intl. Symp. on Parallel and Distributed Processing}, 2008.

\bibitem{Naik2012}
M.~Naik, H.~Yang, G.~Castelnuovo, and M.~Sagiv.
\newblock {Abstractions from Tests}.
\newblock {\em SIGPLAN Not.}, 47(1), Jan. 2012.

\bibitem{Nethercote:2007}
N.~Nethercote and J.~Seward.
\newblock {Valgrind: A Framework for Heavyweight Dynamic Binary
  Instrumentation}.
\newblock {\em SIGPLAN Not.}, 42(6), Jun. 2007.

\bibitem{Omre2008}
W.~Omre.
\newblock {Debug and Trace for Multicore SoCs}.
\newblock Technical report, ARM, 2008.

\bibitem{Rico2011}
A.~Rico, A.~Duran, F.~Cabarcas, Y.~Etsion, A.~Ramirez, and M.~Valero.
\newblock {Trace-driven Simulation of Multithreaded Applications}.
\newblock In {\em Performance Analysis of Systems and Software (ISPASS), 2011
  IEEE International Symposium on}, pages 87--96, 2011.

\bibitem{Ruiz2008}
A.~Ruiz-Alvarez and K.~Hazelwood.
\newblock {Evaluating the Impact of Dynamic Binary Translation Systems on
  Hardware Cache Performance}.
\newblock In {\em IEEE Intl. Symp. on Workload Characterization (IISWC)}, 2008.

\bibitem{Serrano}
M.~Serrano and X.~Zhuang.
\newblock {Building Approximate Calling Context from Partial Call Traces}.
\newblock In {\em Proc. of the 7th Annual IEEE/ACM Intl. Symp. on Code
  Generation and Optimization (CGO)}, 2009.

\bibitem{Simon2008}
B.~Simon, D.~Bouvier, T.-Y. Chen, G.~Lewandowski, R.~McCartney, and K.~Sanders.
\newblock {Common Sense Computing (Episode 4): Debugging}.
\newblock {\em Computer Science Education}, 18(2), 2008.

\bibitem{Srivastava2004}
A.~Srivastava and A.~Eustace.
\newblock {ATOM: A System for Building Customized Program Analysis Tools}.
\newblock {\em SIGPLAN Not.}, 39, 1994.

\bibitem{Uh2007}
G.-R. Uh, R.~Cohn, B.~Yadavalli, R.~Peri, and R.~Ayyagari.
\newblock {Analyzing Dynamic Binary Instrumentation Overhead}.
\newblock 2007.

\bibitem{Upton2011}
D.~Upton and K.~Hazelwood.
\newblock {Finding Cool Code: An Analysis of Source-level Causes of Temperature
  Effects}.
\newblock In {\em Performance Analysis of Systems and Software (ISPASS), 2011
  IEEE International Symposium on}, pages 117--118, 2011.

\bibitem{Upton2009}
D.~Upton, K.~Hazelwood, R.~Cohn, and G.~Lueck.
\newblock {Improving Instrumentation Speed via Buffering}.
\newblock In {\em Proc. of the Workshop on Binary Instrumentation and
  Applications (WBIA)}, 2009.

\bibitem{Wallace2007}
S.~Wallace and K.~Hazelwood.
\newblock {SuperPin: Parallelizing Dynamic Instrumentation for Real-Time
  Performance}.
\newblock In {\em Intl. Symp. on Code Generation and Optimization (CGO)}, Mar.
  2007.

\bibitem{Yu2011}
Z.~Yu, W.~Zhang, and X.~Tu.
\newblock {MT-Profiler: A Parallel Dynamic Analysis Framework Based on
  Two-stage Sampling}.
\newblock In {\em Proc. of the 9th Intl. Conf. on Advanced Parallel Processing
  Technologies (APPT)}, 2011.

\bibitem{Zhao2010}
Q.~Zhao, I.~Cutcutache, and W.-F. Wong.
\newblock {PiPA: Pipelined Profiling and Analysis on Multicore Systems}.
\newblock {\em ACM Trans. Archit. Code Optim.}, 7(3), Dec. 2010.

\end{thebibliography}

\end{document}